\documentstyle[preprint,aps,epsfig]{revtex}
\tightenlines

\begin{document}

\begin{center}
{\Large\bf Matching the observed cosmological constant\\
 with vacuum energy density in AdS}

\bigskip
\bigskip

Zhe Chang\footnote{Email: changz@mail.ihep.ac.cn.}  and
Xin-Bing Huang\footnote{Email: huangxb@mail.ihep.ac.cn.}\\
\medskip
{\em Institute of High Energy Physics, Academia Sinica\\
P. O. Box 918(4), 100039 Beijing, China}

\end{center}

\bigskip
\bigskip
\bigskip

\centerline {\large Abstract}

\bigskip

 We calculate the vacuum energy density
by taking account of
different massive scalar fields in AdS spacetime. It is
found that the mass spectrum of a scalar field in AdS spacetime is
discrete because of a natural boundary condition. The results
match well with the observed cosmological constant.

\vspace{2cm}


\newpage

\section{Introduction}

The cosmological constant has evoked much controversy in both
astronomy and particle physics communities\cite{Weinberg,Carroll92}.
Recent observations of high-redshift supernovae seem to suggest that the
global geometry of the universe may be affected by a positive cosmological
constant\cite{Zehavi}. And all kinds of cosmological
observations, such as Cosmic Microwave Background radiation\cite{Bernardis,Jaffe},
redshifts of the
supernovae and quasars\cite{Perlmutter,Riess98,Riess00}, give a very tiny vacuum energy
density
as $10^{-48} {\rm GeV}^4$~\cite{Carroll00}.

In particle physics, the vacuum is used to mean the
ground state of quantum fields. A relativistic field may be thought of as
a collection of harmonic oscillators of all possible frequencies, and each
possible mode devotes $\frac{1}{2}\hbar\omega$ energy to the vacuum. In
this way, particle physicists\cite{Weinberg} get a huge vacuum energy density as
$2\times10^{71}{\rm GeV}^4$, which is over $120$
orders of magnitude in excess of the value allowed by cosmological
observations. It is
a more challenging problem to explain why the cosmological constant is so
small but non-zero, than to build theoretical models where it exactly
vanishes\cite{Elizalde}.

About twenty years ago, a number of authors discovered that Anti-de Sitter
(AdS) spacetime
generically arose as ground state in supergravity theory, which at the
time
was considered to be among the most promising candidates for quantum
gravity\cite{Abbott,Breitenlohner,Hawking}. The interest on AdS spacetime was
revived by a conjectured duality between string theory in the bulk of AdS and
conformally invariant field theory (CFT) living on the boundary of
AdS\cite{Maldacena}. The AdS/CFT correspondence gives an explicit
relation between Yang-Mills theory and string
theory\cite{Witten,Gubser}. More recently, there has been a renewed interest
in AdS spacetime since progresses in theories of extra dimensions
present us with the enticing possibility to explain some long-standing
particle physics problem by geometrical means\cite{Randall,Arkani}.

The theoretical approaches to the cosmological constant problem can
briefly be classified into three categories:
(1)anthropic principle,
(2)the ``quintessence''\cite{Garriga},
(3)fundamental partical
physics and basic spacetime topology. In many scenarios, the mean
value of the vacuum energy is positive, and
the vacuum energy is related to the gravitational
potential\cite{Elizalde}, or the fluctuation of the vacuum\cite{Gurzadyan}, or the
wormholes which are
described by string quantum cosmology.

It is no
doubt that these studies are fruitful and helpful to make more
progress in understanding the cosmological constant problem. However, it
seems far away from achieving a natural and self-consistent explanation,
which can be checked directly by the update astronomical observations.

In this
Letter, we calculate the energy density of vacuum by taking account of
different massive scalar fields in AdS spacetime. It is
found that the mass spectrum of a scalar field in AdS spacetime is
discrete because of a natural boundary condition\cite{Li,Avis}. The results
match well with the observed cosmological constant.

\section{Equations of motion}

AdS spacetime can be described as a submanifold of a pseudo-Euclidean
five-dimensional embedding space with Cartesian coordinates $\xi^a$ and
metric $g_{\mu\nu}={\rm diag}(1,-1,-1,-1,1)$,

\begin{eqnarray}
(\xi^0)^2-(\xi^1)^2-(\xi^2)^2-(\xi^3)^2+(\xi^5)^2&=&-\frac{1}{\lambda}~,
\nonumber\\
ds^2=(d\xi^0)^2-(d\xi^1)^2-(d\xi^2)^2-(d\xi^3)^2&+&(d\xi^5)^2~,
\end{eqnarray}
where $\lambda$ is a constant ($\lambda<0$). It is obvious that the symmetry
group of AdS is the conformal group $SO(3,2)$. Hence, we can define the
5-dimensional angular momentum of a free particle in AdS,

\begin{equation}
\begin{array}{l}
\displaystyle
L^{\mu\nu}=m_{0}\left(\xi^{\mu}\frac{d\xi^{\nu}}{ds}-
\xi^{\nu}\frac{d\xi^{\mu}}{ds}\right)~,
\end{array}
\end{equation}
the commutation
relations
satisfied by the ten infinitesimal generators are

\begin{eqnarray}
[L_{\alpha\beta},L_{\mu\nu}]=-i(g_{\alpha\mu}L_{\beta\nu}+g_{\beta\nu}
L_{\alpha\mu}-g_{\alpha\nu}L_{\beta\mu}-g_{\beta\mu}L_{\alpha\nu})~&,&
\nonumber\\
\alpha,\beta,\mu,\nu=0,1,2,3,5~&.&
\end{eqnarray}
It is convenient to introduce the Beltrami coordinates $\{x^i\}~(i=0,1,2,3)$ as

\begin{eqnarray}
\sqrt{-\lambda}x^{i}&=&(\xi^5)^{-1}\xi^{i}~.
\end{eqnarray}
In terms of the Beltrami coordinates, AdS is of the form

\begin{eqnarray}
\sigma(x)= 1- \lambda \eta_{ij} x^{i}x^{j}>0~,  \quad
\eta_{ij}={\rm diag}(1,-1,-1,-1)~.
\end{eqnarray}
The Beltrami metric can be deduced directly

\begin{eqnarray}
ds^2&=&(\eta_{ij}\sigma^{-1}+ \lambda \eta_{ir} \eta_{js} x^{r}
x^{s}\sigma^{-2})dx^{i} dx^{j}~.
\end{eqnarray}
In this coordinates, ten group generators $L_{\mu\nu}$ are simply
classified into two categories, $L_{ij}$ correspond to the generators
of the rotation group $SO(3)$, $P_{k}$ generate parallel displacements.
Thus, $L_{ij}$ and $P_{k}$ can be respectively defined as
the angular momentum and the momentum 4-vector in the
Beltrami coordinates

\begin{equation}
\begin{array}{l}
\displaystyle
P^{k}=\frac{1}{\sqrt{-\lambda}}L^{5k}=m_{0}{\sigma^{-1}}
\frac{dx^{i}}{ds}~,
\nonumber
\\
L^{ij}=m_{0} \left( x^{i}P^{j}-x^{j}P^{i} \right)=
m_{0} {\sigma^{-1}} \left(x^{i}\frac{dx^{j}}{ds}-
x^{j}\frac{dx^{i}}{ds} \right)~.
\end{array}
\end{equation}
Einstein's mass formula in AdS is of the form (with $\hbar=c=1$)

\begin{eqnarray}\label{conserve}
m_{0}^{2}&=&\frac{\vert \lambda \vert}{2}L^{\mu\nu}L_{\mu\nu}
= E^{2}- \vec{P}^{2}+\lambda\vec{L}^{2}~,
\nonumber
\\E&=&P^{0}~,\quad \quad \vec{P}=(P^{1},P^{2},P^{3})~,
\end{eqnarray}
The Beltrami metric is invariant under the coordinate transformations

\begin{eqnarray} \label{transformation}
x^{i}\longrightarrow \bar{x}^{i}=\sigma^{1/2}(a)(1-\lambda\eta_{rs}a^{r}
x^{s})^{-1}(x^{j}-a^{j})D^{i}_{j}~,\nonumber\\
D^{i}_{j}=L^{i}_{j} + \lambda\eta_{kl}a^{l}
a^{i}\left[\sigma(a)+\sigma^{1/2}(a)\right]^{-1}L^{k}_{j}~,\\
(L^{i}_{j})_{i,j=0,1,2,3}\in SO(3,1)~,\quad a^{i}~ {\rm are~constants~.}
\nonumber
\end{eqnarray}
In the coordinate $(\xi^0,x^\alpha)$, the $SO(3,2)$ invariant metric
can be written as

\begin{equation}\label{eq5}
  ds^2=\frac{1}{1+\lambda\xi^0\xi^0}d\xi^0d\xi^0-(1+\lambda\xi^0\xi^0)
  \frac{d{\bf x}(I+\lambda{\bf x}'{\bf x})^{-1}d{\bf x}'}
      {1+\lambda{\bf x}{\bf x}'}~,
\end{equation}
where the vector ${\bf x}$ denotes $(x^1,~x^2,~x^3)$ and ${\bf x}'$
the transpose of the vector ${\bf x}$. \\
In the spherical coordinate $(x^1,~x^2,~x^3)\longrightarrow
(\rho,~\theta,~\phi)$, the $SO(3,2)$ invariant metric (\ref{eq5}) is
of the form

\begin{equation}
  ds^2 = \frac{1}{1+\lambda\xi^0\xi^0}d\xi^0d\xi^0
  -(1+\lambda\xi^0\xi^0)\left[\frac{d\rho^2}{(1+\lambda\rho^2)^2}
     +\frac{\rho^2}{(1+\lambda\rho^2)}\left(d\theta^2+\sin^2\theta
d\phi^2\right)\right] ~.
\end{equation}
The Penrose diagram of
AdS shows that there is a horizon\cite{Chang}
in the coordinate $(\xi^0,{\bf x})$.
We limit us in the region of
$\vert\xi^0\vert<\frac{1}{\sqrt{-\lambda}}$.
In this region of AdS, we
can introduce a timelike variable $\tau$ as

\begin{equation}
\sqrt{-\lambda}\xi^0\equiv \sin(\sqrt{-\lambda}\tau)~.
\end{equation}
Then we have a Robertson-Walker-like metric

\begin{eqnarray}
ds^2=d\tau^2-R^2(\tau)[(1+\lambda\rho^2)^{-2}d\rho^2+(1+\lambda\rho^2)^{-1}
\rho^2(d\theta^2+\sin^2\theta d\phi^2)]~,
\end{eqnarray}
where we have used the notation $R(\tau)=\cos(\sqrt{-\lambda}\tau)$.\\
The equation of motion for a massive scalar field in AdS spacetime
 is of the form

\begin{equation}\label{KG}
\begin{array}{l}
\displaystyle\left[\frac{1}{R^3}\frac{\partial}{\partial\tau}
\left(R^3\frac{\partial}
{\partial\tau}\right)-\frac{(1+\lambda\rho^2)^2}{R^2\rho^2}
\frac{\partial}{\partial\rho}\left(\rho^2\frac{\partial}{\partial
\rho}\right)\right.\\[1cm]
\displaystyle~~~~~~~~~~~\left.-\frac{1+\lambda\rho^2}{R^2\rho^2}
\left(\frac{1}{\sin\theta}
\frac{\partial}
{\partial\theta}\left(\sin\theta\frac{\partial}{\partial\theta}\right)-
\frac{1}{\sin^2\theta}\frac{\partial^2}{\partial\phi^2}\right)
+m_{0}^2\right]\Phi(\tau;\rho,\theta,\phi)=0~.
\end{array}
\end{equation}

\section{ Discrete mass spectrum}

We can solve the equation of motion, which was obtained in the last
section, for a massive scalar field\cite{Li,Fronsdal74}  by writing

\begin{eqnarray}
\Phi(\tau;\rho,\theta,\phi)&=&T(\tau)U(\rho)Y_{lm}(\theta,\phi)~.
\end{eqnarray}
The reduced equations of motion in terms of $T(\tau)$, $U(\rho)$ and
$Y_{lm}(\theta,\phi)$ are of the form

\begin{eqnarray}\label{three}
\frac{\partial^2 U}{{\partial\rho}^2}+\frac{2}{\rho}\frac{\partial
U}{\partial\rho}+\frac{k^2}{(1+\lambda\rho^2)^2}U-\frac{l(l+1)}{\rho^2
(1+\lambda\rho^2)}U =0~,\nonumber\\
R^2\frac{d^2T}{{d\tau}^2}+3R\frac{dR}{d\tau}\frac{dT}{d\tau}
+(m_{0}^2R^2+k^2)T = 0~,\\
\frac{\partial^2Y_{lm}}{{\partial\theta}^2}+ {\rm ctg}\theta\frac
{\partial Y_{lm}}{\partial\theta}
+\frac{1}{{\rm sin}^2\theta}\frac{\partial^2Y_{lm}}
{{\partial\phi}^2}+l(l+1)Y_{lm}=0~.\nonumber
\end{eqnarray}
It is obvious that the solutions of the angular part of equation of motion
are the spherical harmonic functions $Y_{lm}(\theta,\phi)$.

The radial function determines the evolution of the model universe.
For convenience, setting $\lambda=-a^{-2}$ and
$\varrho=a^{-1}\rho$, we can rewrite the radial equation as

\begin{eqnarray}
\frac{d^2U(\varrho)}{{d\varrho}^2}+\frac{2}{\varrho}\frac{dU(\varrho)}
{d\varrho}+\left[\frac{k^2a^2}{(1-\varrho^2)^2}-\frac{l(l+1)}{\varrho^2(1-
\varrho^2)}\right]U(\varrho)&=&0~.
\end{eqnarray}
It is obvious that, at $\varrho=0,~\pm{1}$, the radial function is singular. One
can set $U(\varrho)$ as following

\begin{eqnarray}
U(\varrho)&=&\varrho^{l}(1-\varrho^2)^{\mu/2}F(\varrho)~,
\end{eqnarray}
where $\mu$ is a solution of the index equation  $\mu(\mu-2)+k^2a^2=0$.
$F(\varrho)$ is a solution of the hypergeometric equation

\begin{eqnarray}
(1-\varrho^2)\frac{d^2F}{{d\varrho}^2}+\left[\frac{2(l+1)}{\varrho}-
2(l+\mu+1)\varrho\right]\frac{dF}{d\varrho}+\left[\frac{1}{4}-\left(
\mu+l+\frac{1}{2}\right)^2\right]F=0~.
\end{eqnarray}
Therefore, we get the radial function of the form

\begin{eqnarray}\label{radialfunction}
\nonumber
U(\rho)=&C&(\frac{\rho}{a})^{l}\left(1-\frac{\rho^2}{a^2}\right)^
{\frac{1}{2}+\frac{1}{2} \sqrt{1-k^2a^2}} \\
 &\times&_{2}{F_{1}\left(\frac{1}{2}(l+\sqrt{1-k^2a^2}+2),
\frac{1}{2}(l+\sqrt{1-k^2a^2}+1),l+\frac{3}{2};\frac{\rho^2}{a^2}\right)}~,
\end{eqnarray}
where $C$ is the normalization constant.

In terms of the variable $\zeta(\equiv\sin\frac{\tau}{a})$, the timelike
evolution function $T$ satisfies the equation

\begin{eqnarray}\label{time}
(1-\zeta\zeta)\frac{d^2T}{d\zeta d\zeta}-4\zeta\frac{dT}{d\zeta}+
\left(a^2m_{0}^2+\frac{a^2k^2}{1-\zeta\zeta}\right)T&=&0~.
\end{eqnarray}
By introducing $T(\zeta)=(1-\zeta\zeta)^{-\frac{1}{2}}P(\zeta)$, we transform
the timelike evolution equation as the standard associated Legendre
equation

\begin{eqnarray}
(1-\zeta\zeta)\frac{d^2P}{d\zeta d\zeta}-2\zeta\frac{dP}{d\zeta}+
\left(a^2m_{0}^2+2-
\frac{1-a^2k^2}{1-\zeta\zeta}\right)P&=&0~.
\end{eqnarray}
Therefore, the solutions of the timelike evolution equation can be
presented as

\begin{eqnarray}
T_{1}(\tau)&\propto&\frac{1}{\cos\frac{\tau}{a}}
P_{-\frac{1}{2}+\sqrt{\frac{1}{4}+m_{0}^2a^2+2}}^{\sqrt{1-k^2a^2}}
\left(\sin\frac{\tau}{a}\right)~,\nonumber\\
T_{2}(\tau)&\propto&\frac{1}{\cos\frac{\tau}{a}}
Q_{-\frac{1}{2}+\sqrt{\frac{1}{4}+m_{0}^2a^2+2}}^{\sqrt{1-k^2a^2}}
\left(\sin\frac{\tau}{a}\right)~,
\end{eqnarray}
where $P_I^N(\zeta)$ and $Q_I^N(\zeta)$ are associated Legendre functions.
Because $Q_I^N(\zeta)$ become infinite on the boundary
$\vert\xi^{0}\vert=\frac{1}{\sqrt{-\lambda}}$~,
we would ignore $Q_I^N(\zeta)$. 
The natural boundary condition of $P_I^N(\zeta)$
on $\vert\xi^{0}\vert=\frac{1}{\sqrt{-\lambda}}$
 requires $I,~ N$ to be integers.
This gives the discrete mass spectrum of scalar fields in AdS spacetime

\begin{equation}\label{mass}
\begin{array}{l}
a^2m_{0}^2+2=I(I+1)~,\\[0.4cm]
-k^2a^2+1=N^2~,\quad  |N|\leq{I}~.
\end{array}
\end{equation}
The wave functions of
scalar fields in AdS spacetime can be written as following

\begin{equation}\label{solution}
\begin{array}{l}
\Phi_{NIlm}(\tau;\rho,\theta,\phi)\propto U_{Nl}(\rho){(\cos
\frac{\tau}{a})}^{-1} P_I^N\left(\sin \frac{\tau}{a}\right)Y_{lm}(\theta,\phi)~.
\end{array}
\end{equation}
In order to get a scalar field theory on AdS spacetime, we now construct a
Hilbert space by defining the inner product on the whole AdS manifold

\begin{equation}\label{inner}
(\Phi_{N'I'l'm'},\Phi_{NIlm})=\int_{V}^{}\Phi_{N'I'l'm'}^{*}
\Phi_{NIlm}R^{3}(\tau)(1+\lambda\rho^{2})^{-2}{\rho}^{2}
\sin\theta d\tau d\rho d\theta d\phi~.
\end{equation}
Because of the
orthogonality of the associated Legendre functions $P_{I}^{N}(\zeta)$ and
the spherical harmonic functions $Y_{lm}(\theta,\phi)$,
these $\Phi_{NIlm}(\tau;\rho,\theta,\phi)$ form a complete orthonormal basis of
the Hilbert space

\begin{eqnarray}
(\Phi_{N'I'l'm'},\Phi_{NIlm})=\delta_{N'N}\delta_{I'I}\delta_{l'l}\delta_{m'm}~.
\end{eqnarray}

\section{Cosmological constant}

During the last three decades, several approaches of quantizing fields
 in AdS had been acquired.
Since AdS is a homogeneous space of the conformal group $SO(3,2)$ , Fronsdal
 adopted a group-theoretic approach to get canonical quantization of
fields\cite{Fronsdal74,Fronsdal65,Fronsdal75a,Fronsdal75b}. The
covariant self-consistent quantization scheme was devised by
considering the information variance through AdS timelike spatial
infinity\cite{Avis}. For our aim in this paper, we just treat the
AdS timelike spatial infinity as a natural boundary constraint to
obtain the discrete Hamiltonian eigenvalues. In this case, a
Hermitian free scalar field operator is defined by

\begin {eqnarray}
\Phi (x) = \sum_{N,I,l,m} \left(\Phi_{NIlm}(x) {\hat {a}}_{NIlm} + \Phi_{NIlm}
^{\ast}(x)
 {\hat{a}}_{NIlm}^{+}\right)~,
\end{eqnarray}
where the sign $x$ denotes the Beltrami coordinates
 $(x^0,x^1,x^2,x^3)$,
and the operators ${\hat{a}}_{NIlm}$, ${\hat{a}}_{N'I'l'm'}^{+}$ satisfy

\begin{eqnarray}
\lbrack {\hat{a}}_{NIlm},{\hat{a}}_{N'I'l'm'} \rbrack &=&
\lbrack {\hat{a}}_{NIlm}^{+},{\hat{a}}_{N'I'l'm'}^{+} \rbrack = 0~,
\nonumber\\
\lbrack {\hat{a}}_{NIlm},{\hat{a}}_{N'I'l'm'}^{+} \rbrack &=& ~\delta_{NN'}
\delta_{II'}\delta_{ll'}\delta_{mm'}~,\\
{\hat{a}}_{NIlm} ~\vert 0 \rangle &=& 0~~,~~
\vert N,I,l,m \rangle = {\hat{a}}_{NIlm}^{+} \vert 0 \rangle~,
\nonumber
\end{eqnarray}
here $\vert 0 \rangle$ denotes the vacuum state. Thus, the ${\hat{a}}_{NIlm}$ and
${\hat{a}}_{NIlm}^{+}$ are the annihilation and creation operators
on the Fock space, which is
constructed as an infinite tensor product of the simple-harmonic-oscillator
Hilbert spaces.
In quantum field theory, the vacuum is used to mean the ground state
of quantum fields. We can reasonably require that the ground states are
spherically symmetric, {\em i.e.}, $l=m=0$\cite{Gurzadyan}.
Hence, the vacuum energy is simply the sum of ground
  energy of the states $\vert NI00 \rangle$. The parameter
$k$ in the Eq.(\ref{time}) is a wave vector because the timelike
evolution equation should reduce to the evolution equation in Minkowskian spacetime when
$\vert \lambda \vert \longrightarrow 0$. Now we have an upper limitation on
the quantum number
$N$  $(N^2=0,~1)$. According to (\ref{conserve}), the ground state energy of a simple
harmonic oscillator is

\begin{equation}
\begin{array}{rcl}
E^{2}&=&m_{0}^{2}+k^{2}+ \lambda l(l+1) \\
&=&m_{0}^{2}+k^{2}~.
\end{array}
\end{equation}
In the case of continuous wave
vector spectrum in Minkowskian spacetime, summing the ground state energy
$\vert \frac{1}{2}\hbar\omega \rangle$ up to a wave number cutoff $\Lambda
\gg m_{0}$ yields a vacuum energy density,

\begin{eqnarray}\label{density}
\langle \rho \rangle =\frac{1}{2}\int_0^{\Lambda}\frac{4\pi
k^2dk}{(2\pi)^3}\sqrt{m_{0}^2+k^2}
\simeq{\frac{\Lambda^4}{16\pi^2}}~~.
\end{eqnarray}
If we believe general relativity up to the Planck scale\cite{Weinberg},
then we might
take $\Lambda\simeq {(8\pi G)^{-\frac{1}{2}}}$, which would give

\begin{eqnarray}
\langle \rho \rangle
\simeq{2^{-10}\pi^{-4}G^{-2}=2\times{10^{71}}{\rm GeV}^{4}}~~.
\end {eqnarray}
It shows that the energy density of vacuum got by 
the quantum field theory
in Minkowskian spacetime is over $120$ orders of magnitude 
in excess of the value of astronomical
observations\cite{Bernardis,Perlmutter,Riess98}.

As shown in the previous section, the natural boundary
conditions at the points $\xi^0=\pm \frac{1}{\sqrt{-\lambda}}$
assure that the wavevector $k$ be discrete. 
Therefore, the integration
in the formula (\ref{density}) should be alternated 
by summation in this discrete energy
Fock space. In fact, any scalar field of arbitrary mass 
$m_{0}$ gives contributions to the energy density of
vacuum. Because the mass spectrum in the quantum field theory 
on Minkowskian spacetime is continuous, 
it is difficult to get a sum over
different mass modes. Now a discrete mass spectrum 
has been obtained for scalar fields in AdS spacetime, 
thus we can sum the contributions
of all scalar fields with different mass. 
We, therefore, get the energy density of AdS vacuum

\begin{eqnarray}
\langle\rho\rangle&=&2\pi\sum_{m_{0}}\sum_{k}\frac{k^2}{{(2\pi)}^3}\delta
k\sqrt{m_{0}^2+k^2}~,
\end{eqnarray}
where $\delta k=\vert k(N=1)-k(N=0)\vert = \frac{1}{a}$  is the wavevector difference
of two eigen states. Equation (\ref{mass}) is used to get the energy
density of vacuum as following

\begin{eqnarray}
\langle\rho\rangle=\frac{1}{{(2\pi)}^2}\sum_{I=1}^{I_{\rm max}}
\sum_{N=0}^{1} \frac{1-N^2}{a^4}
\sqrt{I(I+1)-N^2-1}~,
\end{eqnarray}
where $I_{\rm max}$ is the cutoff of mass spectrum. We
would estimate $I_{\rm max}$ as the Planck scale,
$E_{\rm Planck}\approx 10^{19}$GeV, which is widely accepted as a
point where conventional field theory breaks down due to quantum
gravitational effects. The maximal energy $E_{\rm max}$ will be the Planck
energy corresponding to the cutoff $I_{\rm max}$,

\begin{eqnarray}
E_{\rm max}=E_{\rm Planck}=\frac{1}{a}\sqrt{I_{\rm max}(I_{\rm max}
+1)-N^2-1}~~,\quad N=0~~.
\end{eqnarray}
In terms of the Planck energy, we obtain a relation of the energy
density of vacuum with the radius $a$ of AdS spacetime

\begin{eqnarray}\label{cosm}
\langle\rho\rangle=\frac{1}{8{\pi}^2}\frac{(E_{\rm Planck})^2}{a^2}~.
\end{eqnarray}
The radius of AdS spacetime can be determined by the observed Hubble 
constant

\begin{eqnarray}
a=\frac{1}{H_{0}}~.
\end{eqnarray}
Thus, we get the energy density of vacuum by the Equation (\ref{cosm})

\begin{eqnarray}
\langle \rho \rangle =\frac{1}{8{\pi}^2}\frac{H^2_0}{G}
=3.63 \times {10^{-48} {\rm GeV}^{4}}~,
\end {eqnarray}
which is in good agreement with the present astronomical observational
value\cite{Zehavi,Jaffe,Carroll00}.
Furthermore, we acquire the cosmological constant

\begin{eqnarray}
\Omega_{\Lambda}=\frac{8\pi G \langle \rho \rangle}{3 H^2_0}
\approx \frac{1}{3\pi}~.
\end{eqnarray}
What we obtain here is in agreement with the hints of the
redshift observations of supernovae and quasars.
References\cite{Bernardis,Guo80,Guo82}  gave an upper limitation of the
curvature $\lambda $ from the estimation of the universe
age and the Hubble constant: $|\lambda|\simeq{10^{-56}}~{\rm cm}^{-2}$.
Our result is also in agreement with this upper limitation for the radius of
AdS spacetime got by different ways.
\section{Concluding remarks}

Since de Sitter found the de Sitter solution of Einstein's equation in
1917, de Sitter and anti-de Sitter spacetime has been studied extensively
by physicists and astronomers. Recent developments in AdS physics include
the AdS/CFT correspondence and theory of extra dimensions.
In this paper, we try to give a new understanding  to the long standing
cosmological constant problem. We assumed that
the topological structure of the whole universe is AdS spacetime. This is consistent
with the Randall-Sundrum model, where a slice of five-dimensional AdS was
used\cite{Randall}.
We got a Robertson-Walker-like metric
which keeps the spacelike submanifold of AdS invariant under
the spacelike subgroup transformation $SO(3)$. Equations of motion for massive scalar fields
were solved exactly by variables separating method.
Solutions indicate that the mass spectrum of scalar fields is
discrete and possible normal modes of scalar fields are limited. These facts
tell us clearly that we can sum the zero-point energies of all kinds
of  scalar fields with different mass.  At last,  an intrinsic relation
between the energy density of vacuum and the curvature of AdS spacetime was
obtained, and the cosmological constant was calculated by using $c/H_0$ for
the radius of AdS spacetime, which matches well with the observational
cosmological constant. It should be pointed out that our results is not
dependent explicitly on dimensions of spacetime\cite{Schmidbuber}. Only for
definiteness, we presented the formalism for AdS$_4$.

\bigskip\bigskip

\centerline{ACKNOWLEDGMENTS}

One of us (Z.C.) would like to thank H. Y. Guo for enlightening discussions.
This work was supported in part by the National Science Foundation of China.

\end{document}